\documentclass[prl, twocolumn, preprintnumbers,amsmath,amssymb,superscriptaddress]{revtex4}
\usepackage{graphicx}
\usepackage{dcolumn}
\usepackage{bm}
\usepackage{subfigure}
\usepackage{multirow}
\usepackage{amsmath}
\usepackage{mathrsfs}
\usepackage{mathtools}
\usepackage{comment}
\usepackage{algpseudocode}
\usepackage{relsize}
\usepackage{color}
\usepackage{amsmath}

 \makeatletter
\newcommand*{\rom}[1]{\expandafter\@slowromancap\romannumeral #1@}
\makeatother
\begin{document}

\title{Entropy-driven physical amplification in multivalent biosensing}
\author{Xiuyang Xia}\email{xiaxy@sustech.edu.cn}
\thanks{These authors contributed equally.}
 \affiliation{
    School of Chemistry, Chemical Engineering and Biotechnology,
Nanyang Technological University, 62 Nanyang Drive, 637459, Singapore
}
 \affiliation{Department of Physics, Southern University of Science and Technology, Shenzhen, Guangdong 518055, China}
  \affiliation{Center for Complex Flows and Soft Matter Research, Southern University of Science and Technology, Shenzhen, Guangdong 518055, China}
 \affiliation{Arnold Sommerfeld Center for Theoretical Physics and Center for NanoScience, Department of Physics,
    Ludwig-Maximilians-Universit\"at M\"unchen, Theresienstraße 37, D-80333 M\"unchen, Germany}
\author{Yuhan Peng}
\thanks{These authors contributed equally.}
 \affiliation{
    School of Chemistry, Chemical Engineering and Biotechnology,
Nanyang Technological University, 62 Nanyang Drive, 637459, Singapore
}

\author{Ran Ni}\email{r.ni@ntu.edu.sg}
 \affiliation{
    School of Chemistry, Chemical Engineering and Biotechnology,
Nanyang Technological University, 62 Nanyang Drive, 637459, Singapore
}


\begin{abstract}
Sensitive detection of low-abundance molecular targets is widely assumed to require enzymatic amplification, such as PCR, to achieve low detection limits. In amplification-free platforms, sensitivity is traditionally constrained by equilibrium binding affinity. Here we show that multivalent linker entropy provides a distinct physical route to exponential sensitivity enhancement in purely equilibrium sensing architectures. Using a statistical-mechanical theory supported by grand canonical Monte Carlo simulations, we demonstrate that redistributing a fixed total interaction strength over increasing linker valency exponentially lowers adsorption thresholds. This scaling emerges not from stronger energetic affinity, but from the rapid growth of combinatorial binding configurations, revealing entropy as an intrinsic amplification mechanism. Consequently, detection limits can be tuned independently of bond strength, enabling ultrasensitive responses without enzymatic replication. Our results establish a general physical design principle for engineering amplification-free detection systems capable of approaching PCR-level sensitivities through entropy-driven collective effects.
\end{abstract}

\maketitle

\section{Introduction}
Sensitive detection of biomarkers and pathogens requires achieving low detection limits at physiologically relevant analyte concentrations. In amplification-based platforms such as PCR, extraordinary sensitivity is obtained through exponential molecular replication~\cite{pcr1,pcr2}. In contrast, amplification-free biosensing systems must rely directly on binding thermodynamics and collective molecular interactions to generate detectable signals~\cite{art1,art2}. Understanding how such systems can approach amplification-level sensitivity without enzymatic replication remains a fundamental physical challenge~\cite{art2}.


Multivalent interactions provide a general physical mechanism for generating nonlinear and highly cooperative responses in systems composed of many weak binding units~\cite{kitov2003nature,huskens2006multivalent,martinez2011designing}.
By distributing interactions over multiple binding sites, multivalency allows individually weak bonds to collectively stabilize macroscopic states, leading to sharp transitions in adsorption, adhesion, and molecular recognition~\cite{mulval1}.
In the context of surface adsorption, statistical-mechanical frameworks have shown that multivalent binding can produce superlinear, and even superselective, responses to changes in environmental parameters~\cite{martinez2011designing}.
Importantly, existing studies have primarily focused on how multivalency sharpens response slopes, enabling discrimination based on receptor density or target abundance, whereas much less is known about how multivalency controls the location of response thresholds themselves, particularly under constraints on total interaction strength.

In many experimentally relevant systems, multivalent recognition does not occur through direct adsorption alone, but instead proceeds via linker-mediated architectures, in which soluble molecules bridge ligands on one object to receptors on another.
Such bridging motifs are ubiquitous across bioanalytical and soft-matter platforms, including antibody-based sandwich immunoassays~\cite{engvall1971enzyme,voller1976enzyme}, DNA-mediated nanoparticle assays~\cite{mirkin1996dna,taton2000scanometric,cao2002nanoparticles} and whole-genome detection~\cite{xu2023whole,xia2024designing}.
From a physical perspective, linker-mediated interactions fundamentally alter the structure of multivalent binding by introducing an additional degree of freedom: linkers can form productive bridges connecting the two surfaces, or remain non-productively sequestered by dangling from either side.
This competition between bridging and sequestration gives rise to response behaviors that differ qualitatively from direct multivalent adsorption, including non-monotonic responses characterized by a low-concentration ``turn-on'' threshold and a high-concentration ``turn-off'' threshold that together define a finite operating window~\cite{tate2004interferences,voller1976enzyme}.

These features raise a fundamental question concerning multivalent interacting systems under constrained interaction strength.
Specifically, when the \emph{total} binding strength between two objects is fixed, what controls the position of the adsorption threshold?
Is the threshold set primarily by energetic affinity, or can collective and entropic effects associated with multivalency qualitatively reshape it?
In linker-mediated architectures, this question becomes particularly acute because linker valency provides a natural means of redistributing a fixed interaction strength over multiple binding arms, necessarily weakening individual bonds while expanding the space of accessible binding configurations, thereby introducing an intrinsic trade-off between energetic stabilization and combinatorial entropy.

Here we address this question by developing a minimal statistical-mechanical framework for linker-mediated multivalent adsorption.
Combining a Langmuir-like adsorption description~\cite{martinez2011designing,xia2024designing}, a theory for multivalent bridging~\cite{xia2025designed}, and grand canonical Monte Carlo simulations, we systematically isolate the role of linker valency under the constraint of fixed total binding strength.
We find that redistributing a fixed interaction strength over an increasing number of linker arms can exponentially shift the adsorption threshold to lower linker concentrations.
This strong and counterintuitive scaling does not originate from enhanced energetic affinity, but instead from the rapid growth of the combinatorial entropy associated with multivalent binding configurations, revealing an entropy-dominated mechanism for threshold positioning.

\begin{figure}[t!]
    \centering
    \includegraphics[width=0.45\textwidth]{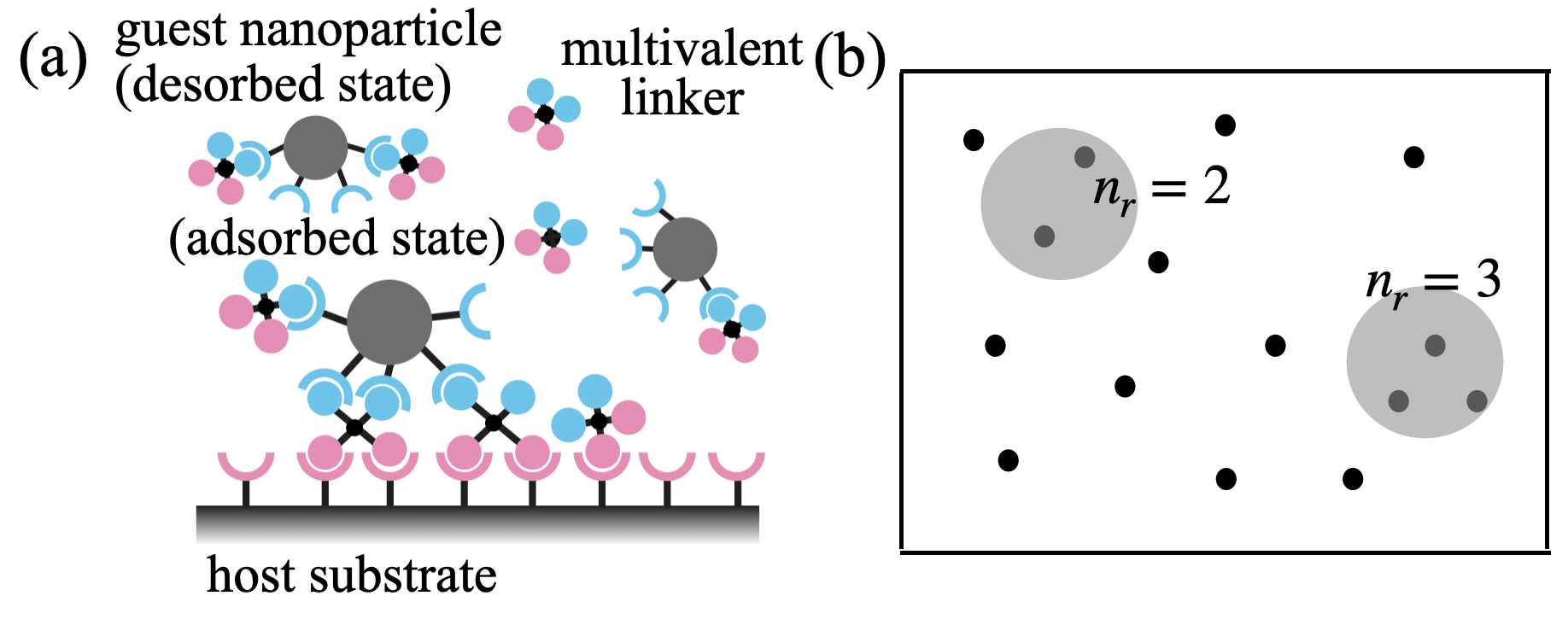}
    \caption{
    \textbf{Multivalent linker-mediated adsorption of guest particles.}
    (a) Schematic illustration of the multivalent linker-mediated interaction between guest nanoparticles and the host substrate. Guest particles that form bridges to the substrate via linkers are considered in the adsorbed state, while those without such bridges are in the desorbed state.
    (b) Schematic of the GCMC simulation setup. Black dots represent immobile receptors randomly distributed on the substrate. Large gray circles indicate guest particles, each of which can interact only with receptors located within its own shaded region via linkers.
    }
    \label{fig1}
\end{figure}
\section{Results}

\subsection{Model and simulations}
We consider a coarse-grained model where guest particles adsorb onto a host substrate via multivalent linkers whose two distinct ends bind to ligands on the guest particles and receptors on the host substrate, respectively (Fig.~\ref{fig1}a). 
Following a Langmuir-like framework~\cite{martinez2011designing}, we consider a substrate site with $n_r$ immobile receptors that can accommodate at most one guest particle.  
Thus, each site exists either in a \emph{desorbed} state, where no linker bridge is formed, or in an \emph{adsorbed} state, where at least one linker bridges between the guest particle and receptors on the site.
A guest particle is coated with $n_l$ mobile ligands, where the mobility of ligands allows them to potentially bridge with any receptor at its corresponding site~\cite{xia2024designing}. 
Linkers are treated as an ideal gas of density $\rho = e^{\beta\mu}/\Lambda^3$, with the chemical potential $\mu$, de Broglie wavelength $\Lambda$, and the inverse thermal energy $\beta = 1/k_B T$, where $k_B$ and $T$ are the Boltzmann constant and system temperature, respectively. 
Each linker possesses $\kappa_l$ and $\kappa_r$ binding ends for ligands and the receptors, with individual binding free energies $f_l$ and $f_r$, respectively. The formation of a bridging linker is penalized by a mean-field configurational free energy cost $f_{\rm cnf}$. 

We formulate the grand potential of the system in both adsorbed and desorbed states. Our approach is based on the framework in Ref.~\cite{xia2025designed}, which has been validated by thermodynamic integration for systems with explicit multivalent linkers. The central idea is to treat each linker as interacting only with receptors on a single site and ligands on the associated guest particle, assuming that the linker size is small compared to the site and guest particles. Thus, linkers mediate an effective interaction between the host and the guest, without coupling with other sites.
In the adsorbed state, besides free linkers in solution, each linker may dangle from a receptor, dangle from a ligand, or bridge between receptors and ligands, thereby connecting the substrate and the particle.
By applying the saddle-point approximation, we determine the equilibrium number of linkers in each state: $M_l$ for those dangling on ligands, $M_r$ for those dangling on receptors, and $Q$ for those that form bridges. The resulting grand potential for the adsorbed state is~\cite{sup_info}
\begin{equation}\label{eq:F_att}
\beta F_{\rm ad} = \sum_{i \in \{l, r\}} \left( n_i \log \bar{n}_i - \bar{n}_i - M_i \right) - Q,
\end{equation}
where $\bar{n}_i$ denotes the number of unbound ligands ($i = l$) or receptors ($i = r$) in the adsorbed state.
Similarly, for the desorbed state, only dangling linkers exist, as no bridges can form when the guest particle is absent. The grand-canonical potential in the desorbed state is given by
\begin{equation}\label{eq:F_de}
\beta F_{\rm de} = \sum_{i \in \{l, r\}} \left( n_i \log \bar{n}_i' - \bar{n}_i' - M_i' \right),
\end{equation}
where $M_l'$ and $M_r'$ represent the numbers of linkers dangling from ligands and receptors, respectively, and $\bar{n}_i'$ represents the number of unbound ligands or receptors when the guest particle is infinitely far from the site. $M$, $Q$, and $\bar{n}$ are numerically calculated by solving a set of self-consistent equations, in which the convergence of the solution for arbitrary linker valency is ensured~\cite{xia2025designed}.

We assume that adsorption events on individual substrate sites are independent, with no interaction between particles adsorbed on adjacent sites. For each site, we take the desorbed state as the reference, and define the referenced partition function of the adsorbed state as $q = e^{-\beta (F_{\rm ad} - F_{\rm de})} - 1$~\cite{angioletti2017exploiting}. The averaged grand partition function of all sites is then given by $\Xi = \langle 1 + z_g q \rangle_{\langle n_r \rangle}$, where $z_g$ is the activity of guest particles in the bulk solution, proportional to the particle density and adsorption volume~\cite{martinez2011designing,xia2024designing}. $\langle \cdot \rangle_{\langle n_r \rangle}$ calculates the average over the Poisson distribution of uncorrelated, immobile receptors, characterized by the mean $\langle n_r \rangle$.
The probability of nanoparticle adsorption follows a Langmuir-type isotherm:
\begin{equation}\label{theory_pred}
\theta := \frac{\partial \log \Xi}{\partial \log z_g} = \left\langle \frac{z_g q}{1 + z_g q} \right\rangle_{\langle n_r \rangle} .
\end{equation}

To validate our theoretical prediction and provide a parameter mapping for future studies, we performed Grand Canonical Monte Carlo (GCMC) simulations of guest nanoparticles in contact with the host substrate coated with explicit receptors. Due to the low density and the multivalent nature of guest particles in most experiments, the equilibration time for 3D simulations is extremely long. Therefore, we simulate a monolayer of guest particles on the substrate where the guest particles can be modelled as 2D hard disks, and they are in chemical equilibrium with a bulk 3D reservoir above. The adsorption and desorption of guest particles are treated as the insertion and deletion of particles from the reservoir in the simulation~\cite{sup_info}.
Guest particles are modeled as hard disks with diameter $\sigma = 1$ on a two-dimensional substrate in chemical equilibrium with a reservoir at chemical potential $\mu_g$~\cite{xia2023role,xia2024designing} (see Fig.~\ref{fig1}b). The activity of guest particles is given by $z_g = e^{\beta \mu_g} \pi \sigma^2 h_0 / (4 \Lambda^3)$, where $h_0=1$ is the thickness of the adsorption layer. Immobile receptors are randomly distributed on the substrate following a Poisson distribution with number density $\rho_r$. Mobile ligands on each guest particle can interact with receptors located within the area underneath the particle. 
Guest particles move on a two-dimensional plane of box length $L$ with periodic boundary conditions and interact via hard-disk repulsion. The maximum number of adsorption sites is set to $N_{\rm max} = 4L^2 / (\pi \sigma^2)$, and the adsorption probability is calculated as $\theta = \langle N_g \rangle / N_{\rm max}$, where $\langle N_g \rangle$ is the average number of adsorbed guest particles. 

We focus on how linker multivalency modulates two key features of the adsorption systems: sensitivity and selectivity. 
Here, sensitivity refers to the lowest detectable value of a given environmental parameter $\chi^\ast$ at which adsorption becomes significant. Operationally, we define the sensitivity threshold $\chi^\ast$ by the condition $\theta(\chi^\ast) = 0.1$.  
Selectivity quantifies the degree, to which adsorption responds to changes in $\chi$, and is characterized by the selectivity parameter $\alpha_\chi = {\mathrm{d} \log \theta}/{\mathrm{d} \log \chi}$. When $\alpha_\chi > 1$, the adsorption increases superlinearly with respect to $\chi$ with $\theta \sim \chi^{\alpha_\chi}$. Conversely, $\alpha_\chi < -1$ indicates that the desorption also proceeds in a superlinear fashion. Both situations are referred to as $\chi$-dependent superselectivity.

\begin{figure}[t!]
    \centering
    \includegraphics[width=0.49\textwidth]{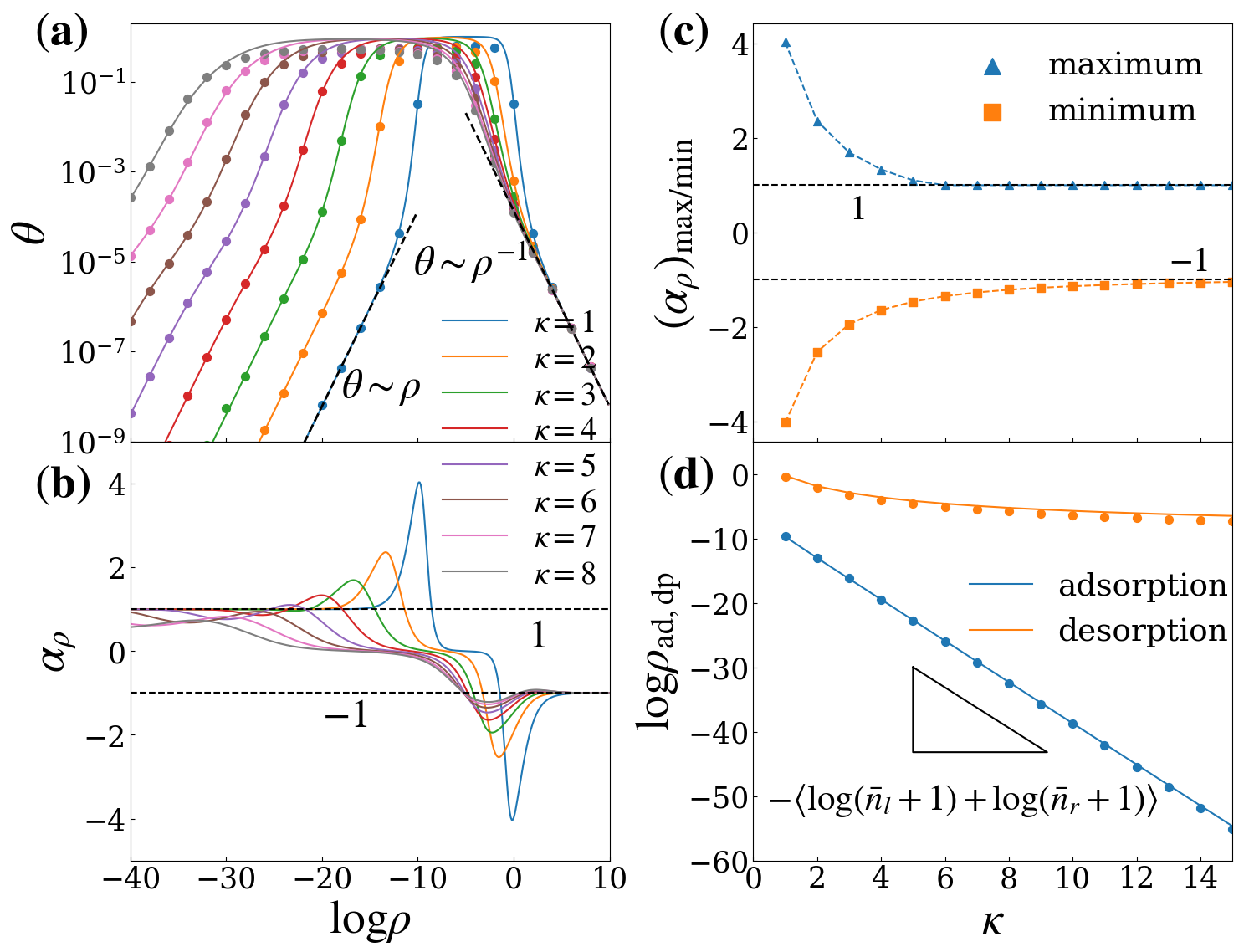}
    \caption{\textbf{$\rho$-dependent superselectivity and ultra-sensitivity.} Here, $\kappa \beta f_l=\kappa \beta f_r=-5$, $n_l=\langle n_r \rangle=10$, $\beta f_{\rm cnf}=2$, and $z_g=10^{-5}$. The symbols are simulation results, and the solid curves are the theoretical prediction (Eq.~\ref{theory_pred}). (a) The adsorption probability $\theta$ as a function of $\mathrm{log} \rho$ under different valencies of linkers. The dashed lines indicate the low-concentration and high-concentration limits of $\theta$~\cite{sup_info}. (b) The selectivity $\alpha_\rho$ to linker concentration as a function of $\mathrm{log} \rho$ under different valencies from theoretical prediction~\cite{sup_info}. (c) The maximal and minimal selectivity $(\alpha_\rho)_{\mathrm{max/min}}$ as a function of linker valency $\kappa$ from theoretical prediction. The dashed lines are guides of eyes where the curves converge to (d) Log of the adsorption and desorption $\mathrm{log} \rho_{\rm ad,dp}$ as a function of linker valency $\kappa$. The small triangle shows the converging slope of the desorption curve (Eq.~\ref{slope}).}
    \label{fig2}
\end{figure}

\subsection{Ultrasensitive $\rho$-response}
To investigate how linker valency $\kappa$ affects the adsorption of guest nanoparticles, we perform both theoretical analysis and GCMC simulations, considering symmetric linkers with $\kappa = \kappa_l = \kappa_r$. We fix the overall binding strength to isolate the effect of linker valency so that any change in the adsorption of nanoparticles arises solely from the combinatorial entropy associated with multivalency. Accordingly, in all calculations we fix the total binding strength as $f_{\rm tot} = \kappa f_l = \kappa f_r$.

As shown in Fig.~\ref{fig2}a, the adsorption $\theta$ increases linearly with $\rho$ at low concentrations~\cite{sup_info}, followed by a pronounced superlinear regime. Remarkably, at high linker concentrations, after a saturation plateau, $\theta$ begins to decrease as $\rho$ increases further, which is a distinct feature of linker-mediated interactions\cite{xia2020scienceadv,Rogers2019}. The corresponding selectivity parameter $\alpha_\rho$, shown in Fig.~\ref{fig2}b, reveals that for small and moderate linker valency $\kappa$, both adsorption and desorption transitions are superselective, i.e., the maxima and minima of $\alpha_\rho$ exceed 1 in magnitude ($\alpha_{\rho,\max} > 1$, $\alpha_{\rho,\min} < -1$).
Interestingly, as $\kappa$ increases, the selectivity weakens: $\alpha_{\rho,\max}$ decreases and $\alpha_{\rho,\min}$ increases, both approaching $1$ and $-1$ in the large valency limit (Fig.\ref{fig2}c). This is because, as the linker valency increases, fewer linker molecules are required to bridge, and eventually a single linker may suffice to fully bridge the host and guest, so that the superselectivity with respect to linker concentration vanishes.
In the limit of zero particle activity $z_g \to 0$ and strong individual binding $f_{l/r} \to -\infty$, our mean-field analysis predicts that the maximal selectivity decays as $\alpha_{\rho,\max} \approx \max[(\min(n_l, n_r))/\kappa$,1]~\cite{sup_info}, indicating that the number of bridges ultimately limits the attainable superselectivity. When $\kappa$ exceeds $\min(n_l, n_r)$, the system loses its superselective response to linker concentration. In realistic regimes of finite $f_{l/r}$ and $z_g$, the selectivity is reduced even further, consistent with MC simulations.

In addition to selectivity, linker valency also strongly influences the adsorption and desorption sensitivity thresholds, $\rho_{\rm ad}$ and $\rho_{\rm dp}$. As shown in Fig.~\ref{fig2}d, $\rho_{\rm ad}$ decreases exponentially with increasing $\kappa$, whereas $\rho_{\rm dp}$ decreases and then reaches a plateau at large $\kappa$. Since the overall binding strength is fixed, these trends arise from the growing combinatorial entropy introduced by linker multivalency. To elucidate the underlying physics, we examine the large-$\kappa$ limit at a single site. 

At the adsorption sensitivity threshold $\rho_{\rm ad}$, both adsorbed and bridging linkers are extremely sparse, because only a very small number of linkers are sufficient to fully bind the guest particle to the site. Thus, the effective interaction between a guest particle and its site mainly depends on the number of unbound ligands and receptors~\cite{sup_info}
\begin{equation}
\beta F_{\rm eff} \simeq \sum_{i = l, r} \left[ n_i \log \left( \frac{\bar{n}_i}{n_i} \right) - (\bar{n}_i - n_i) \right].
\label{eq:F_eff_large_kappa}
\end{equation}
Under the Poisson distribution of receptor numbers, this yields the adsorption sensitivity threshold~\cite{sup_info}
\begin{equation}\label{slope}
    \log \rho_{\rm ad} \sim  -\lim_{\kappa \to +\infty}\left[\mathrm{log}\left(\bar{n}_l +1 \right) + \mathrm{log}\left(\bar{n}_r +1 \right) \right] \kappa.
\end{equation}
This shows that $\rho_{\rm ad}$ decreases exponentially with increasing linker valency $\kappa$, with the exponent given by the constant $-\left[ \log(\bar{n}_l + 1) + \log(\bar{n}_r + 1)\right]$ at the large $\kappa$ limit, which agrees quantitatively with GCMC simulations (Fig.~\ref{fig2}d).
Thus, as the total binding strength $\kappa f_{l/r}$ is fixed, this exponential scaling reflects the dominant contribution of the combinatorial entropy associated with linker multivalency. 
This intriguing result suggests that even with a fixed overall binding strength, increasing $\kappa$ can exponentially lower the detection limit, thereby enabling ultrasensitive detection.

Different from the effect of valency $\kappa$ on the adsorption of nanoparticles, the desorption threshold $\rho_{\rm dp}$ decreases with increasing $\kappa$ and reaches a plateau at large $\kappa$ limit (Fig.~\ref{fig2}d).
This is because all receptors and ligands are fully occupied at large $\kappa$. Further increasing $\kappa$ does not increase the number of bound ends per linker, thus has no effect on the adsorption probability $\theta$ and corresponding $\rho_{\rm dp}$~\cite{sup_info}. This behavior is qualitatively consistent with the purely entropy-driven phase behavior observed in programmable colloidal atom-electron equivalents at large linker valency~\cite{xia2025designed}.

\subsection{Detection among non-specific binders}
\begin{figure*}[t]
    \centering
    \includegraphics[width=0.99\textwidth]{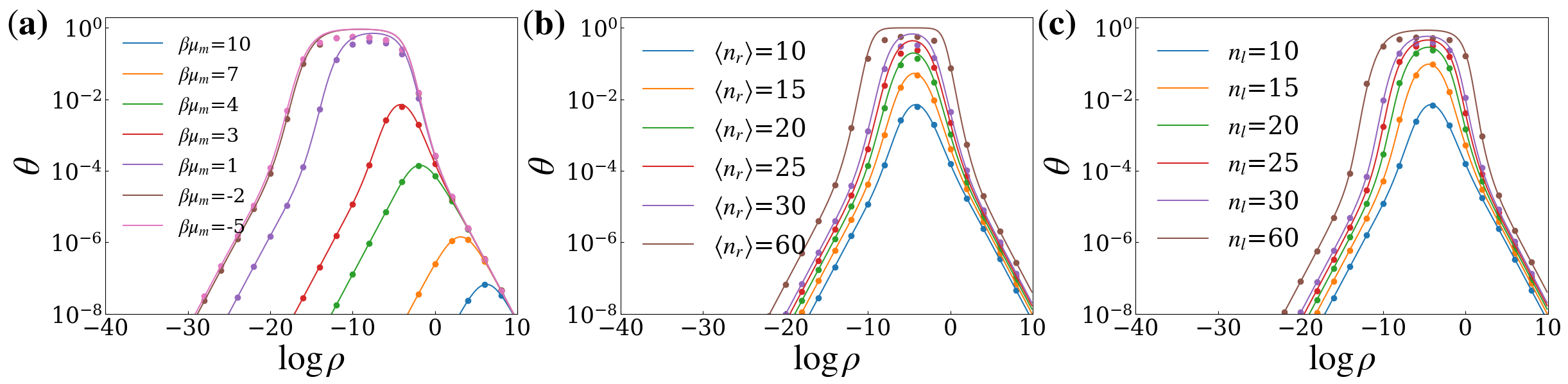}
    \caption{\textbf{Influence of non-specific binders.} Here, $\kappa \beta f_l=\kappa \beta f_r=-5$, $\kappa=3$, $\beta f_m=1$, $\beta f_{\rm cnf}=2$, and $z_g=10^{-5}$. The symbols are simulation results, while the solid curves are the theoretical prediction (Eq.~\ref{theory_pred}). (a) The adsorption probability $\theta$ as a function of $\log \rho$ for different non-specific binder chemical potential $\mu_m$, where $n_l=\langle n_r \rangle=10$. (b) The adsorption probability $\theta$ as a function of $\log \rho$ under different average receptor numbers $n_r$ with $n_l = 10$ and $\beta \mu_m=3$. (c) The adsorption probability $\theta$ as a function of $\log \rho$ under different nanoparticle valencies $n_l$ with $\langle n_r \rangle = 10$ and $\beta \mu_m=3$.}
    \label{fig3}
\end{figure*}

In complex environments such as biological fluids or environmental samples, numerous molecules other than the targeted linkers are present~\cite{shrivastav2021comprehensive,sampad2024label,sangkham2021review}. These untargeted species can adsorb onto sensor surfaces or recognition elements with weak non-specific interactions. Generally, such non-specific binding would introduce a background signal and reduce the selectivity and sensitivity of the detection system. Hence, careful consideration and mitigation of the environmental non-specific binding is essential for developing a reliable detection system in real-world applications. 
To this end, we generalize our model by considering non-specific binders. Those non-specific species are modeled as weak monovalent binders that can bond weakly with either ligands or receptors but cannot form bridges between them~\cite{binders}. Here we only consider the non-specific binders, as they are the most common environmental noise in bio-detections, and the untargeted molecules bridging between ligands and receptors are rare~\cite{binders2}. Those binders are present in the environment with individual non-specific binding free energies $f_m$ and treated as an ideal gas of density $\rho_m=e^{\beta \mu_m}/\Lambda^3$ with the chemical potential $\mu_m$.

We formulate the grand potential of the system with the mixture of the target linkers and non-specific binders by applying the saddle-point approximation. The resulting grand potential for the adsorbed state is~\cite{sup_info}
\begin{equation}
    \beta F_{\rm ad} =\sum_{i \in \{ l,r\}} \left( n_i  \log \bar{n}_i -\bar{n}_i -M_i -B_i\right)-Q,
\end{equation}
where $B_i$ is the number of adsorbed binders onto the substrate $(i=r)$ or the particle $(i=l)$. Likewise, for the desorbed state, where no bridge forms, the grand potential is 
\begin{equation}
    \beta F_{\rm dp} =\sum_{i \in \{l,r \}}\left( n_i \log \bar{n}_i'-\bar{n}_i'-M_i'-B_i'\right),
\end{equation}
where $B_i'$ is the number of adsorbed binders when the particle is far from the substrate. 

The theoretical predictions and the simulation results for different chemical potential of binders are shown in Fig.~\ref{fig3}a. One can see that with increasing the chemical potential of the non-specific binders, the adsorption of guest nanoparticles mediated by targeted linkers is significantly disrupted, and the saturation range gradually disappears with the detection threshold $\rho_{\rm ad}$ moving to a larger value, which indicates the vanishing of the ultra-sensitivity. 
This suppression of linker-mediated adsorption is due to the competition between non-specific binders and targeted linkers for binding sites. By adsorbing onto ligands and receptors, the non-specific binders effectively reduce the number of available binding partners for the bridge formation. Therefore, a strategy to mitigate this effect is to increase the total number of ligands or receptors, thereby restoring available binding sites. As shown in Fig.~\ref{fig3}b and c, increasing the densities of receptors grafted on host substrate or the number of ligands coated on guest nanoparticles progressively recovers the detection sensitivity of targeted linkers.

\section{Discussion}
In summary, we have developed a theoretical framework that elucidates how linker multivalency governs the sensitivity and selectivity of bio-detection systems. Our analysis shows that, at a fixed overall binding strength, increasing linker valency exponentially lowers the detection threshold, thereby enabling ultrasensitive readouts even at extremely low target concentrations. 
In practical terms, amplification-free multivalent systems engineered in this manner may substantially narrow the sensitivity gap relative to PCR-based amplification assays. That such performance can arise purely from entropy-driven combinatorial effects, without enzymatic replication, highlights multivalency as a physical analogue of signal amplification. The exponential scaling identified here provides a mechanistic explanation for how ART-type platforms can achieve surprisingly high sensitivity despite lacking molecular amplification.
As a result, even sparse populations of multivalent targets can generate robust adsorption signals, which is highly desirable for early-stage biomarker and pathogen diagnostics.
At the same time, increasing linker valency reduces superselectivity with respect to linker concentration, reflecting a fundamental trade-off between sensitivity and concentration-based discrimination. Furthermore, the presence of non-specific binders can suppress ultrasensitivity when their concentration becomes sufficiently high. This effect can be mitigated by increasing the densities of receptors grafted on host substrate or the number of ligands coated on guest nanoparticles, thereby restoring the detection sensitivity.
Taken together, these findings indicate that optimal detection performance does not necessarily require ligands and receptors with the strongest possible binding affinities. Instead, those capable of forming multiple simultaneous interactions with the target may be more effective. 
These principles are directly applicable to rapid antigen tests, DNA-mediated aggregation assays, and multivalent immunosensors, where signal generation relies on collective binding rather than enzymatic replication.
This principle also rationalizes the recently developed ultrasensitive whole-genome detection strategy based on DNA-functionalized colloids, where long single-stranded DNAs (ssDNAs) act as ``multivalent linkers'', driving colloidal aggregation at extremely low ssDNA concentrations~\cite{xu2023whole}.
Our results provide general design principles for engineering multivalent detection systems and highlight the fundamental role of combinatorial entropy in biomolecular recognition. The theoretical insights presented here offer practical guidelines for the development of next-generation biosensors and diagnostic assays capable of ultra-sensitive and highly specific detection in complex biological environments.

\section{Method}
In this work, grand-canonical Monte Carlo (GCMC) simulations are performed for 2D systems with implicit ligands and linkers. The guest particles are modelled as hard disks with radius $\sigma/2$, each coated with ligands. The ligands can interact with the receptors within $\sigma/2$ to the center of the guest particles. The receptors are distributed on the host substrate according to the Poisson distribution. 

Due to the low density and the multivalent nature of guest particles in most experiments, the equilibration time for 3D simulations is extremely long. Therefore, we simulate the a monolayer of guest particles on the substrate where the guest particles can be modelled as 2D hard disks, and they are in chemical equilibrium with a bulk 3D reservoir above. The adsorption and desorption of guest particles are treated as the insertion and deletion of particles from the reservoir in the simulation. The probabilities of attempting translational trial moves, particle addition trial moves, and deletion trial moves are set to $0.3$, $0.35$, and $0.35$, respectively. 

Moreover, if a particle is inserted at a random position $\bf s'$ and its effective interaction is described by the adsorption partition function $q({\bf s'}) = e^{-\beta [F_{\rm ad}({\bf s'}) - F_{\rm de}({\bf s'})]} - 1$. This trial move is accepted by the probability 
\begin{multline}
    acc\left( N_g \to N_g+1\right)= \\ {\rm min}\left[ 1,\frac{V q({\bf s'})}{\Lambda^2 (N_g+1)}e^{\beta [\mu_g -U({\bf s'},N_g+1)+U({\bf s'},N_g)]}\right],
\end{multline}
where $V$ is the area of the 2D plane, $\Lambda$ is the thermal de Broglie wavelength, $N_g$ is the number of guest particles before the insertion, $\beta \mu_g$ is the chemical potential of the reservoir of guest particles, and $U({\bf s'},N_g)$ quantifies the hard-core repulsion. The deletion of a particle at position $\bf s'$ is accepted with probability 
\begin{multline}
    acc\left( N_g \to N_g-1\right)= \\{\rm min}\left[ 1,\frac{\Lambda^2 N_g}{V q({\bf s'})}e^{-\beta [\mu_g +U({\bf s'},N_g-1)-U({\bf s'},N_g)]}\right],
\end{multline}
where $N_g$ is the number of adsorbed guest particles before deletion.
The translational trial move of a particle from a position $\bf s$ to $\bf s'$ is accepted with a probability
\begin{equation}
    acc\left( \bf s \to s'\right)= {\rm min} \left[ 1, \frac{q({\bf s'})}{q({\bf s})}e^{-\left( \beta U({\bf s'},N_g) -\beta U ({\bf s},N_g)\right)}\right].
\end{equation}
Moreover, periodic boundary conditions are applied in all directions. 
In each simulation, $10^7$ MC moves are performed for equilibration and $ 10^8$ MC moves for production with sampling every $5$ moves. The adsorption probability $\theta$ is computed as $N_g/N_{\rm max}$. 

\section{acknowledgments}
 X. X. acknowledges support from the Alexander von Humboldt-Stiftung.
 This work was financially supported by the Academic Research Fund from the Singapore Ministry of Education (RG151/23, and RG88/25) and the National Research Foundation, Singapore, under its 29th Competitive Research Program (CRP) Call (NRF-CRP29-2022-0002).

\section{Data Availability Statements}
The data that support the findings of this article are openly available~\cite{data_info}. 

\bibliography{abbr}

\end{document}